\documentclass[twoside,reqno]{qts-proc}
\usepackage{epsfig,cite}
\usepackage{amssymb,amsmath}
\usepackage{times}
\begin{document}
\sloppy \raggedbottom
\setcounter{page}{1}

\newpage
\setcounter{figure}{0}
\setcounter{equation}{0}
\setcounter{footnote}{0}
\setcounter{table}{0}
\setcounter{section}{0}



\title{A Generalized Rule For Non-Commuting Operators in Extended Phase Space}

\runningheads{S.Nasiri}{A Generalized Rule For Non-Commuting
Operators...}

\begin{start}


\author{S. Nasiri}{1,2},
\coauthor{S. Khademi}{1}, \coauthor{S. Bahrami}{1}, \coauthor{F.
Taati}{2}

\address{Department of Physics, Zanjan Univ., ZNU, Zanjan, Iran.}{1}
\address{Institute for Advanced Studies in Basic Sciences, IASBS, Zanjan, Iran.}{2}


\begin{Abstract}
A generalized quantum distribution function is introduced. The
corresponding ordering rule for non-commuting operators is given
in terms of a single parameter. The origin of this parameter is in
the extended canonical transformations that guarantees the
equivalence of different distribution functions obtained by
assuming appropriate values for this parameter.
\end{Abstract}
\end{start}


\section{Introduction}
The hamiltonian formulation of classical mechanics treats the
generalized coordinates, $q$, and momenta, $p$, as independent
variables and in a symmetric way. In quantization procedure,
however, one chooses the q- or p-representation destroying the
equal standing of the two variables. In his pioneering work,
Wigner [1] proposed quantum distribution function in phase space
keeping the symmetry of $q$ and $p$ and defined the expectation
values of observables in the manner of classical statistical
mechanics. Here, we use the extended phase space technique
proposed by Sobouti and Nasiri [2] and the canonical
transformations in this space to introduced a generalized quantum
distribution function. It is known that for any distribution
function there is an ordering rule of non-commuting operators.
Thus, for generalized distribution function a generalized ordering
rule is introduced, as well. Assuming appropriate values for the
parameter involved, different distribution functions and their
corresponding ordering rules can be obtained. The origin of these
parameters is in the extended canonical transformations in the
extended phase space.

The layout of the paper is as follows: In section 2, a brief
review of the extended phase space formalism is presented. In
section 3, the generalized distribution function is introduced.
Section 4, is devoted to conclusions.

\section{A Review of EPS formalism}

A direct approach to quantum statistical mechanics is proposed by
Sobouti and Nasiri [2], by extending the conventional phase space
and by applying the canonical quantization procedure to the
extended quantities in this space. Here, a brief review of this
formalism is presented. For more details, the interested reader
may consult Sobouti and Nasiri [2].

Let ${\cal L}^q(q,\dot{q})$ be a lagrangian specifying a system in
$q$ space. A trajectory of the system in this space is obtained by
solving the Euler-Lagrange equations for $q(t)$,
\begin{equation}
\frac{d}{dt}\frac{\partial {\cal L}^q}{\partial \dot
q}-\frac{\partial {\cal L}^q}{\partial q}=0.
\end{equation}
The derivative $\frac{\partial {\cal L}^q}{\partial \dot q}$
calculated on an actual trajectory, that is, on a solution of Eq.
(1), is the momentum $p$ conjugate to $q$ . The same derivative
calculated on a virtual orbit, not a solution of Eq. (1), exists.
It may not, however, be interpreted as a canonical momentum. Let
$H(p,q)$ be a function in phase space which is the hamiltonian of
the system, whenever $p$ and $q$ are canonical pairs. It is
related to ${\cal L}^q$ through the Legendre transformation,

\begin{equation}
{H}(\frac{\partial {\cal L}^q}{\partial \dot
q},q)=\dot{q}\frac{\partial {\cal L}^q}{\partial \dot q}-{\cal
L}^q(q,\dot q).
\end{equation}
For a given ${\cal L}^q$, Eq. (2) is an algebraic equation for
$H$. One may, however, take a different point of view. For a given
functional form of $H(p,q)$, Eq. (2) may be considered as a
differential equation for ${\cal L}^q$. Its non unique solutions
differ from one another by total time derivatives. One may also
study the same system in the momentum space. Let ${\cal
L}^p(p,\dot p)$ be a lagrangian in $p$ space. It is related to
$H(p,q)$ as follows,

\begin{equation}
H(p,\frac{\partial {\cal L}^p}{\partial \dot
p})=-\dot{p}\frac{\partial {\cal L}^p}{\partial \dot p}+{\cal
L}^p(p,\dot p).
\end{equation}
Here the functional dependence of $H$ on its argument is the same
as in Eq. (2). In principle, Eq. (3) should be solvable for ${\cal
L}^p$ up to an additive total time derivative term. Once ${\cal
L}^p$ is known the actual trajectories in $p$ space are obtainable
from an Euler-Lagrange equation analogous to Eq. (1) in which  $q$
is replaced by $p$. That is
\begin{equation}
\frac{d}{dt}\frac{\partial {\cal L}^p}{\partial \dot
p}-\frac{\partial {\cal L}^p}{\partial p}=0.
\end{equation}
The derivative $\frac{\partial {\cal L}^p}{\partial \dot p}$ along
an actual $p$ trajectory is the canonical coordinate conjugate to
$p$ . Calculated on a virtual
orbit, it is not.\\

A formulation of quantum statistical mechanics is possible by
extending the conventional phase space and by applying the
canonical quantization procedure to the extended quantities in
this space. Assuming the phase space coordinates $p$ and $q$ to be
independent variables on the virtual trajectories, allows one to
define momenta $\pi_{p}$ and $\pi_{q}$, conjugate to $p$ and $q$,
respectively. One may combine the two pictures and define an
extended lagrangian in the phase space as the sum of $p$ and $q$
lagrangians,

\begin{equation}
{\cal L}(p,q,{\dot p},{\dot q})=-{\dot p}q-{\dot q}p + {\cal L}^{p}(p,{\dot p})+ {\cal L}^{q}(q,{\dot q}).
\end{equation}
The first two terms in Eq. (5) constitute a total time derivative.
The equations of motion are

\begin{equation}
\frac{d}{dt}\frac{\partial {\cal L}}{\partial \dot
p}-\frac{\partial {\cal L}}{\partial p}=\frac{d}{dt}\frac{\partial
{\cal L}^p}{\partial \dot p}-\frac{\partial {\cal L}^p}{\partial
p}=0,
\end{equation}
\begin{equation}
\frac{d}{dt}\frac{\partial {\cal L}}{\partial \dot
q}-\frac{\partial {\cal L}}{\partial q}=\frac{d}{dt}\frac{\partial
{\cal L}^q}{\partial \dot q}-\frac{\partial {\cal L}^q}{\partial
q}=0.
\end{equation}
The $p$ and $q$ in Eqs. (5-7) are not, in general, canonical
pairs. They are so only on actual trajectories and through a
proper choice of the initial values. This gives the freedom of
introducing a second set of canonical momenta for both p and q.
One does this through the extended lagrangian. Thus

\begin{equation}
\pi_{p} = \frac{\partial {\cal  L} }{\partial {\dot p} } = \frac{\partial {\cal L}^p}{\partial {\dot p}}-q,
\end{equation}

\begin{equation}
\pi_{q} = \frac{\partial {\cal  L }}{\partial {\dot q} } = \frac{\partial  {\cal L} ^q}{\partial {\dot q} }-p.
\end{equation}
Evidently, $\pi_p$ and $\pi_q$ vanish on actual trajectories and
remain non zero on virtual ones. From these extended momenta, one
defines an extended hamiltonian,
\begin{eqnarray}
{\cal H}(\pi_p ,\pi_q , p,q) & = & {\dot p}\pi_p +{\dot q}\pi_q -{\cal L}= H(p+\pi_q,q)-H(p,q+\pi_p)\nonumber\\
& = & \sum\frac{1}{n!}\left\{\frac{\partial^nH}{\partial p^n}\pi_{q}^n-\frac{\partial^nH}{\partial q^n}\pi_{p}^n\right\}.
\end{eqnarray}
Using the canonical quantization rule, the following postulates
are outlined: a) Let $p,q,\pi_p$ and $\pi_q$ be operators in a
Hilbert space, $\bf X$, of all square integrable complex
functions, satisfying the following commutation relations
\begin{equation}
[\pi_{q},q]=-i\hbar,\hspace{2cm}\pi_{q}=-i\hbar\frac{\partial}{\partial q},
\end{equation}
\begin{equation}
[\pi_{p},p]=-i\hbar,\hspace{2cm}\pi_{p}=-i\hbar\frac{\partial}{\partial p},
\end{equation}
\begin{equation}
[p,q]=[\pi_{p},\pi_{q}]=0.
\end{equation}
By virtue of Eqs. (11-13), the extended hamiltonian, ${\cal H}$,
will also be an operator in ${\bf X}$. b) A state function
$\chi(p,q,t)\in{\bf X}$ is assumed to satisfy the following
dynamical equation
\begin{eqnarray}
i\hbar\frac{\partial \chi}{\partial t}&=&{\cal
H}\chi=[H(p-i\hbar\frac{\partial}{\partial q},q)
-H(p,q-i\hbar\frac{\partial}{\partial p})]\chi\nonumber\\
&=&\sum\frac{(-i\hbar)^n}{n!}\left\{\frac{\partial^nH}{\partial
p^n}\frac{\partial^n}{\partial q^n}-\frac{\partial^nH}{\partial
q^n}
\frac{\partial^n}{\partial p^n}\right\}\chi.
\end{eqnarray}
c) The averaging rule for an observable $O(p,q)$, a c-number
operator in this formalism, is given as
\begin{equation}
<O(p,q)>=\int O(p,q)\chi^{*}(p,q,t)dpdq.
\end{equation}
To find the solutions for Eq. (14) one may assume

\begin{equation}
\chi(p,q,t)=F(p,q,t)e^{-ipq/\hbar}
\end{equation}
The phase factor comes out due to the total derivative in the
lagrangian of Eq. (5), -d(pq)/dt. The effect is the appearance of
a phase factor, $exp(-ipq/\hbar )$, in the state function that
would have been in the absence of the total derivative. It is
easily verified that

\begin{equation}
(p-i\hbar\frac{\partial}{\partial q})\chi=i\hbar\frac{\partial
F}{\partial q}e^{-ipz/\hbar}
\end{equation}

\begin{equation}
(q-i\hbar\frac{\partial}{\partial p})\chi=i\hbar\frac{\partial
F}{\partial p}e^{-ipz/\hbar}
\end{equation}
Substituting Eqs. (17) and (18) in Eq. (14) and eliminating the
exponential factor gives

\begin{equation}
{H(-i\hbar\frac{\partial}{\partial q},q)
-H(p,-i\hbar\frac{\partial}{\partial p})}F=i\hbar\frac{\partial
F}{\partial t}.
\end{equation}
Equation (19) has separable solutions of the from

\begin{equation}
F(p,q,t)=\psi(q,t)\phi^{\ast}(p,t),
\end{equation}
where $\psi(q,t)$ and $\phi(p,t)$ are the solutions of the
Schrodinger equation in $q$ and $p$ representations, respectively.
The solution of the form (16) associated with anti-standard
ordering rule satisfies Eq. (14) and is one possible distribution
function. For more details on the admissibility of the
distribution functions, their interesting properties and the
correspondence rules, one may consult Sobouti and Nasiri [2].

\section{Generalized distribution function and ordering rule}
In 1932 Wigner [1] proposed the distribution

\begin{equation}
W(q,p)=\frac{1}{\pi \hbar}\int <q-y|\hat
{\rho}|q+y>e^{ipy/\hbar}dy,
\end{equation}
for a system in mixed state represented by a density matrix
$\hat{\rho}$. The expectation value for an operator $\hat{O}$
calculated with $W(q,p)$ has the same value as ordinary quantum
average with wave function $\psi$, i.e.
\begin{equation}
<\hat{O }>_\psi=\int W(q,p)O(q,p)dpdq,
\end{equation}
where $O(q,p)$ is a classical function corresponding to operator
$\hat O$ and is given according to wigner prescription by[3]
\begin{equation}
O(q,p)=\int<q-\frac{z}{2}|\hat{O}|q+\frac{z}{2}>e^{ipz/\hbar}dz.
\end{equation}
Similarly, the distribution function of Eq. (16) could be written
as
\begin{equation}
\chi(q,p)=\frac{1}{2\pi\hbar}\int<q|\hat{\rho}|q+z>e^{ipz/\hbar}dz,
\end{equation}
and corresponding averaging rule
\begin{equation}
<\hat{O}(q,p)>_{\chi}=\int \chi(q,p)O(q,p)dpdq,
\end{equation}
where the classical function $O(q,p)$ is given by
\begin{equation}
O(q,p)=\int<q|\hat{O}|q+z>e^{ipz/\hbar}dz.
\end{equation}
Let a generalized distribution function be defined as
\begin{equation}
P_\alpha(q,p)=\frac{1}{2\pi\hbar}\int <q+\alpha
z|\hat{\rho}|q+(\alpha+1)z>e^{ipz/\hbar}dz,
\end{equation}
and the corresponding ordering rule as
\begin{equation}
O_\alpha(q,p)=\int<q+\alpha z|\
\hat{O}|q+(\alpha+1)z>e^{ipz/\hbar}dz,
\end{equation}
where $\alpha$ is a parameter specifying the given distribution
function, that is, for $\alpha=\frac{-1}{2}$ one has the Wigner
function, for $\alpha=0$ one has the standard distribution
function and etc. The origin of $\alpha$ is in the canonical
transformations in the extended phase space that serve to obtain
the different distribution functions [2]. Equations (27)and (28)
may be rewritten as
\begin{equation}
P_\alpha(q,p)=\frac{1}{2\pi\hbar}\int<q|e^{i\hat{p}\alpha
z/\hbar}\hat{\rho}e^{-i\hat{p}\alpha z/\hbar}|q+z>e^{ipz/\hbar}dz,
\end{equation}

\begin{equation}
O_\alpha(q,p)=\int<q|e^{i\hat{p}\alpha z/\hbar}\hat{O}e^{-
i\hat{p}\alpha z/\hbar}|q+z>e^{ipz/\hbar}dz.
\end{equation}
Equations (29) and (30) relate the generalized distribution
function and the corresponding ordering rule to those of the (SN)
representation by a unitary transformation. Thus, using the EPS
representation as the basis, one may introduce different
representations using Eqs. (29) and (30) for different values of
$\alpha$. As an example, consider an operator as
$\hat{O}=\hat{q}^m \hat{p}^n$. Using Eq. (28) the classical
function corresponding to this operator after a lengthy
calculations becomes
\begin{equation}
O_\alpha (p,q)=
\sum^m_{r=0}\left(^m_r\right)\frac{n!}{(n-r)!}(i\hbar\alpha)^rp^{n-r}q^{m-r},
n\leq r.
\end{equation}
Assuming $\alpha=\frac{-1}{2}$, Eq. (31) will be identical with
that of Wyle-Wigner ordering rule [3], and for $\alpha=0$, one
obtains the standard ordering rule [4].

\section{Conclusions}

In the phase space approach to quantum mechanics each distribution
function corresponds to an ordering rule. In this paper we
introduce a generalized distribution function and corresponding
generalized ordering rule. A parameter, $\alpha$ serves to specify
the different quantum distribution functions. The origin of the
parameter $\alpha$ comes from the extended canonical
transformations in the extended phase space. The cases of SN and
Wigner distribution functions are worked out as examples.
%


\section*{Acknowledgments}

The authors wish to thank Prof. Sobouti for his helpful comments.



\begin{thebibliography}{10}



\bibitem{1}Wigner, E. P.,Phys. Rev., 40, 749,
(1932).
\bibitem{2}Sobouti, Y. and Nasiri, S., Int.J.Mod.Phys.B, 18, p.7 (1993).
\bibitem{3}Hillery, M., Oconnel, R. F., Scully, M. O. and Wigner, E. P.,
Phys. Rep. C106, 121, (1984).
\bibitem{4}Sobouti, Y. and Nasiri, S.,Tr. J. Phys., 18, 458, (1995).
\end{thebibliography}
\end{document}